\documentclass[12pt,draftcls,journal,onecolumn,romanappendices]{IEEEtran}

\usepackage{amssymb,amsmath,amsthm}
\usepackage[dvips]{graphicx}
\providecommand{\abs}[1]{\lvert#1\rvert}

\title{Multiuser Diversity in Downlink Channels:
When does the Feedback Cost Outweigh the Spectral Efficiency Gain?}
\author{Amogh Rajanna and Nihar Jindal\\Dept
of ECE, University of Minnesota\\Email:\{raja0088,nihar\}@umn.edu}

\newtheorem{Propi}{Proposition}
\newtheorem{Theoi1}{Theorem}
\newtheorem{Theoi2}[Theoi1]{Theorem}
\newtheorem{Lem}{Lemma}

\begin{document}
\maketitle
\begin{abstract}
In this paper, we perform a cost-benefit analysis of multiuser diversity in single antenna broadcast channels.
It is well known that multiuser diversity can be beneficial but there is a significant cost associated with acquiring instantaneous CSI. We perform a cost-benefit analysis of multiuser diversity for 2 types of CSI feedback methods, dedicated feedback and SNR dependent feedback, quantifying how many users should feedback CSI from a net throughput perspective. Dedicated feedback, in which orthogonal resources are allocated to each user,
has significant feedback cost and this limits the amount of available multiuser diversity that
can be used. SNR dependent feedback method, in which only users with SNR above a threshold attempt to feedback, has relatively much smaller feedback cost and
this allows for all of the available multiuser diversity to be used. Next, we study the effect of single user multiantenna techniques, which reduce the SNR variation, on the number of feedback users neccessary. It is seen that a broadcast channel using single user multiantenna techniques should reduce the number of feedback users with the spatial dimension.
\end{abstract} 

\section{Introduction}
\label{sec:intro}
In broadcast wireless systems, where the channels to different users fade independently, it is well known that multiuser diversity can be beneficial \cite{Knopp}. At any point in time, it is likely that at least one user will have a very good channel realization. If the BS is aware of the user channels, it can schedule data transmission to the user with the best instantaneous channel at a high rate, thereby achieving better performance. The quality of the selected channel increases with the number of users, and a very well known result is that the spectral efficiency increases double logarithmically in the number of users \cite{So}.

Although multiuser diversity can provide significant benefits (multiuser diversity is a key component of contemporary cellular systems UMTS, LTE etc), there is
also a non-negligible cost associated
with obtaining instantaneous channel state information (CSI) at the BS. Such CSI is obtained through explicit feedback of the instantaneous SNR from each of the users or through utilization of uplink pilots 
when the channel is reciprocal. Thus, the cost is in terms of the system resources bandwidth and power used to acquire
CSI and it increases with the number of users. To illustrate why the cost-benefit tradeoff is interesting and non-trivial, note that the standard CSI feedback methods
consume resources that are \textit{linear} in the number of users who feedback, whereas
the downlink spectral efficiency increases only \textit{double-logarithmically} in the number of users. Even if a more efficient method is
used to acquire CSI, another fundamental overhead arises from the fact that the BS must signal which of the users has been selected for
transmission. This overhead is \textit{logarithmic} in the number of users, and thus is also larger, in a scaling sense, than the
benefit of multiuser diversity. Hence, 
an optimized system should appropriately balance the cost and benefit of multiuser diversity, and a basic design question is to determine how many users the
BS should acquire CSI from or in other words,
how much of the available multiuser diversity should be used.

The feedback cost associated with obtaining CSI at the BS depends on the specific CSI feedback method used and the number of feedback users. In this paper, we consider 2 types of CSI feedback. The first method is dedicated feedback, where each user is allocated an orthogonal resource per coherence time i.e., a fixed number of uplink symbols to be used either for SNR feedback or for transmission of uplink pilots. For this method, the feedback resources increase \textit{linearly} with the number of users. The second method considered is SNR dependent feedback \cite{So,Gesbert1,Tang}, where only users who have an instantaneous SNR above a threshold attempt to feedback instead of having every user always feedback its SNR. This method exploits the fact that
only users with strong channels are likely to be scheduled, and thus feedback is not required from users with weak channels.
Since only the users with strong channels feedback their SNR, the feedback must be performed on a shared random access channel (e.g., using slotted
ALOHA or spread spectrum) \cite{Tang, Love}.

In this paper, we consider both the FDD and TDD (reciprocal) systems. In this paper, the FDD and TDD  systems differ in how the uplink and downlink bandwidths are related.
For each system, the cost-benefit tradeoff of multiuser diversity for both the dedicated and SNR dependent feedback methods is quantified in terms of the key system parameters like the average SNR and the blocklength.
Also, wherever possible, the optimal number of feedback users is quantified. 

Although the focus of the paper is on single antenna broadcast channels,
we also study the effect of single user multiantenna techniques on multiuser diversity. More specifically, how the single user multiantenna techniques, which reduce SNR or mutual information (MI) variation, affect the multiuser diversity order that needs to be used.

The remainder of the paper is organized as follows. In section \ref{sec:prob}, the system model and analysis framework are introduced. In sections \ref{sec:FDDproblem} and \ref{sec:TDDproblem}, the cost-benefit analysis of multiuser diversity in the FDD and TDD systems is performed. In section \ref{sec:SU_Multiant}, we study the effect of single user multiantenna techniques on multiuser diversity and we conclude in Section \ref{conc_sec}. 
\section{System Model}
\label{sec:prob}
 We consider a single antenna broadcast channel with $K$ users. A Rayleigh block fading model is assumed. Every coherence block has $\check{T}$ symbols, which is termed the downlink resource blocklength and is given by $\check{T}=W_cT_c$, where $W_c$ is the coherence bandwidth and $T_c$ is the coherence time.
The BS performs single user scheduling once every coherence block and to do so the BS 
collects SNR information from $K$ users at the begining of every coherence block. Through common downlink training from the BS, users will learn their SNR and explicitly feed it back on the uplink. We assume that the SNR value received at the BS is error free and delay free. \footnote{Although practical systems exhibit continuous fading, with possibly different coherence times and bandwidths for different users,
we are able to get to the core insights of this problem by focusing on the simple scenario of block fading with the same block size for all users. Furthermore, the effect of feedback delay can be accounted for by choosing an appropriate effective coherence time.}
After learning the downlink SNR of the $K$ users, the BS selects the user with the largest instantaneous SNR for transmission.\footnote{A key assumption in this paper is that all the users have the same average SNR. In a practical system users will have different average SNR's, in which case the proportionally fair scheduler \cite{David}, which chooses the user with the largest instantaneous SNR relative to its own average SNR, should be used.}

\subsection{Quantifying the benefit of multiuser diversity}
\label{FBBenefit}
In this subsection, we quantify the benefit of multiuser diversity i.e., 
how the spectral efficiency increases w.r.t the number of feedback users when using dedicated feedback. The spectral efficiency is 
\begin{equation}
 C_{df}\left( K \right)=\mathbb{E}\left[\log_2(1+P\gamma_{df})\right]\label{spec_eff}
\end{equation}
where $\gamma_{df}=\max_{1\le i \le
K}\gamma_{i}$ is the scheduled SNR, $\gamma_{i}=\lvert h_i\rvert^2$, $h_i\sim \mathcal{CN}\left(0,1\right)$ is the downlink channel of $i^{th}$ user and $P$ is the average SNR.

From \cite{Jangchen}, the spectral efficiency (\ref{spec_eff}) has the following closed form expression
\begin{align}
 C_{df}\left( K \right)&=\sum_{k=1}^{K}\binom{K}{k}(-1)^{k+1}~e^{\frac{k}{P}}
~E_1\left(\frac{k}{P}\right)\label{exC}
\end{align}
where $E_1(.)$ is the Exponential integral $\displaystyle{E_1(x)=\int^{\infty}_x \frac{e^{-t}}{t}~dt}$.
Although (\ref{exC}) gives a closed form expression, it does not quantify how $C_{df}\left( K \right)$ increases
with $K$. To quantify the growth rate of $C_{df}\left( K \right)$ with $K$, the following upper bound and lower bound on
$C_{df}\left( K \right)$ are useful.

\begin{Propi}
\label{prop1}
 For large $K$, the spectral efficiency $C_{df}\left( K \right)$ is bounded as
  \begin{align}
&\log_2 \left(1+P \left(\log K-\log\log\log K\right)\right)+ o(1)\leq
C_{df}\left( K \right) \leq \log_2 \left(1+P\left(0.58 +
\log  K\right)\right)\label{LUB}
\end{align}
\end{Propi}
\begin{IEEEproof}
 See Appendix \ref{sec:prop1proof}.
\end{IEEEproof}

Based on (\ref{LUB}), the spectral efficiency can be approximated as
\begin{align}
C_{df}\left( K \right)&\approx \tilde{C}_{df}\left( K \right)=\log_2 \left(1+P \log K\right)=O\left(
\log \log
K \right)\label{gKapprox}
\end{align}

\begin{figure}
\centering
\includegraphics[scale=0.7,width=0.7\textwidth]{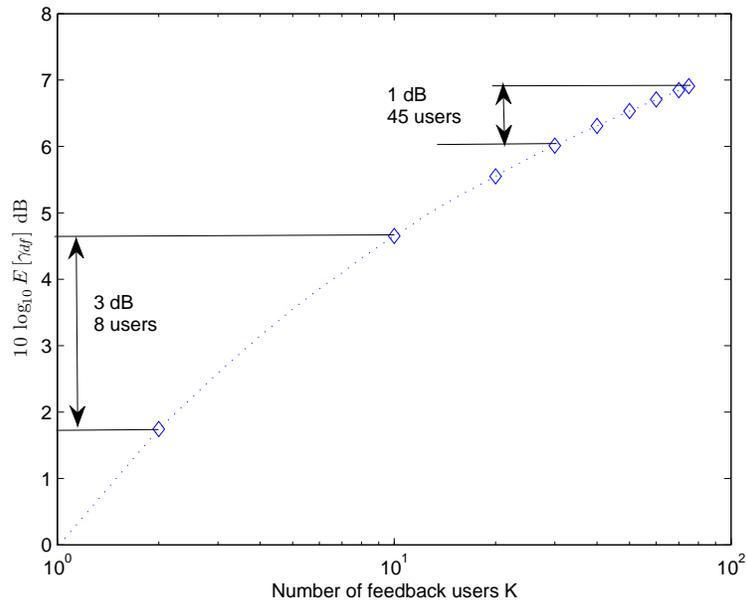}
\caption{Average scheduled SNR in dB scale plotted against the number of feedback users.}
\label{EYkplot_df}
\end{figure}

In Fig.\ref{EYkplot_df}, the average scheduled SNR $\mathbb{E}[\gamma_{df}]$ in dB scale is plotted against the number
of feedback users. The expression for $\mathbb{E}[\gamma_{df}]$ is given in Appendix \ref{sec:prop1proof}. From Fig.\ref{EYkplot_df}, it is seen that the scheduled SNR increases quickly for the first few users, but afterwards the rate of increase slows dramatically.
\subsection{FDD System}
\label{feedbackFDD}
\begin{figure}
\centering
\includegraphics[scale=0.7,width=0.7\textwidth]{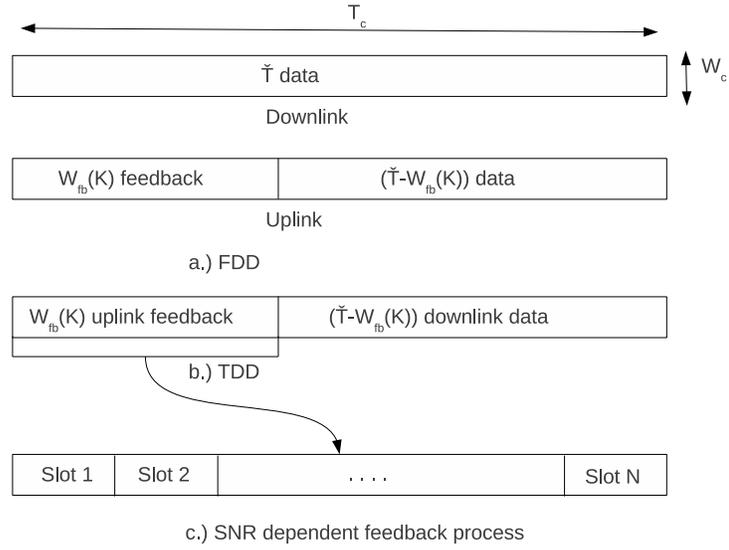}
\caption{a.) Separate uplink and downlink bandwidths in the FDD system. b.) Common bandwidth for uplink and downlink in the TDD system. c.) Feedback bandwith in the SNR dependent feedback process.}
\label{Fig2}
\end{figure}

In Frequency Divison Duplexing (FDD) systems, the uplink and downlink have separate bandwidths. Fig.\ref{Fig2}.a shows the structure of the uplink and downlink bandwidths in a FDD system. The downlink bandwidth is used only for downlink data whereas the uplink bandwidth is used for SNR feedback (to aid downlink) and uplink data. Every coherence block, the uplink bandwidth is split into 2 pieces, the first one used for SNR feedback and the second one used for data transmission. At the begining of each uplink bandwidth, $K$  users feedback SNR to the BS. Let $\mathrm{W_\mathrm{{fb}}(K)}$ be the number of
symbols used for SNR feedback on the uplink. $\mathrm{W_\mathrm{{fb}}(K)}$ will be defined separately for each of the 2 types of SNR
feedback methods in the later sections. The remaining $\check{T}-\mathrm{W_\mathrm{{fb}}(K)}$ symbols are used for uplink data. The uplink rate is determined by the uplink data bandwidth and uplink spectral efficiency. The uplink data bandwidth $\mathrm{W_\mathrm{{data}}(K)}$, normalized to the total uplink bandwidth $\check{T}$, is
\begin{align}
 &\mathrm{W_\mathrm{{data}}(K)}=1-\frac{\mathrm{W_\mathrm{{fb}}(K)}}{\check{T}}
 \label{FDD_data_bw}
\end{align}
We model the uplink channel as an AWGN channel with SNR $P$, and hence the uplink spectral efficiency is 
\begin{align}
C_U=\log_2\left( 1+P \right)
\end{align}
  Therefore, the uplink rate is
\begin{align}
R_{U}\left( K \right)&=\mathrm{W_\mathrm{{data}}(K)}\cdot C_U=\left(1-\frac{\mathrm{W_\mathrm{{fb}}(K)}}{\check{T}}
\right)\log_2\left( 1+P \right)\label{uplinkrate}
\end{align}
Since the downlink bandwidth is only used for data, on downlink $\mathrm{W_\mathrm{data}}(K)=1$. The downlink spectral efficiency is 
\begin{equation}
 C\left( K \right)=\mathbb{E}\left[\log_2(1+P\gamma_{sch})\right]
\end{equation}
where $\gamma_{sch}$ is the scheduled SNR, expression for which depends on the type of SNR feedback method. Therefore,
the downlink rate is
\begin{equation}
 R_{D}\left( K \right)=\mathrm{W_\mathrm{{data}}(K)}\cdot C\left( K \right)=\mathbb{E}\left[\log_2(1+P\gamma_{sch})\right]
\end{equation}
Increasing the number of feedback users increases the downlink rate but decreases the uplink data bandwidth and thus, decreases the uplink rate. \footnote{SNR feedback using downlink bandwidth can be used to increase the spectral efficiency on the uplink. This leads to a very similar tradeoff between uplink and downlink rates, and the basic insights found here apply to that scenario as well.} Clearly there exists a tradeoff between the uplink and downlink rates.
This tradeoff can be quantified by considering the maximization of the weighted sum of the uplink and downlink rates.
\begin{equation}
\max_K \left[\lambda_R R_{D}\left( K \right)+R_{U}\left( K \right)\right] \label{FDDoptim}
\end{equation}
where $\lambda_R$ is the weight factor, specifying the preference of the downlink rate over the uplink rate.
Thus, our fundamental objective is to determine the optimal number of feedback users as function of the system parameters blocklength, average SNR and weight factor.\footnote{If the total number of users in the system is $K_{total}$, our objective is to find the number of feedback users $K$ from among $K_{total}$ which achieves an optimal balance between the uplink and downlink rates.}

\subsection{TDD System}
\label{FBTDD}
Time Divison Duplexing (TDD) systems differ from the FDD systems in that one common bandwidth is used for both the uplink and downlink. Fig.\ref{Fig2}.b shows the structure of the common bandwidth for both the uplink and
downlink in a TDD system. Every coherence block, the bandwidth is split into 2 pieces, the first one used for SNR feedback on the uplink and the second one used for data transmission on the downlink.\footnote{In TDD systems employing dedicated feedback, the users can alternatively send uplink pilots to allow the BS to estimate the per-user SNR. Although we don't explicitly refer to this mode of operation in the paper, it is straightforward to see that it fits into our framework.}
For the sake of simplicity, it is assumed that there is no uplink data. It is explained later in section \ref{sec:TDDproblem} that even if the uplink data is present, it won't change the cost-benefit tradeoff of multiuser diversity in a TDD system.
At the begining of each bandwidth, users feedback SNR to the BS on the uplink using $\mathrm{W_\mathrm{{fb}}(K)}$ symbols. The number of data symbols on the downlink is $\check{T}-\mathrm{W_\mathrm{{fb}}(K)}$.
The downlink rate is determined by the downlink data bandwidth and the spectral efficiency.
It is given by
\begin{align}
 &R_D(K)=\mathrm{W_\mathrm{{data}}(K)}\cdot C\left( K \right)=\left(1-\frac{\mathrm{W_\mathrm{{fb}}(K)}}{\check{T}}
\right)\mathbb{E}\left[\log_2(1+P\gamma_{sch})\right]
\label{DLrate}
\end{align}

From (\ref{DLrate}), we see that the two terms which determine the downlink rate have a tradeoff with the number of
feedback users, since the spectral efficiency increases and the data bandwidth decreases in $K$. Thus, our fundamental objective is to determine the optimal number of feedback users, in terms of maximizing the downlink rate, as function of the system parameters blocklength and average SNR. 
\section{FDD System}
\label{sec:FDDproblem}
In this section, a cost-benefit analysis of multiuser diversity in a FDD system is performed. We first study dedicated feedback and then address SNR-dependent feedback.

\subsection{Dedicated Feedback}
\label{FDD_ded_fb}
\begin{figure}
\centering
\includegraphics[scale=0.7,width=0.7\textwidth]{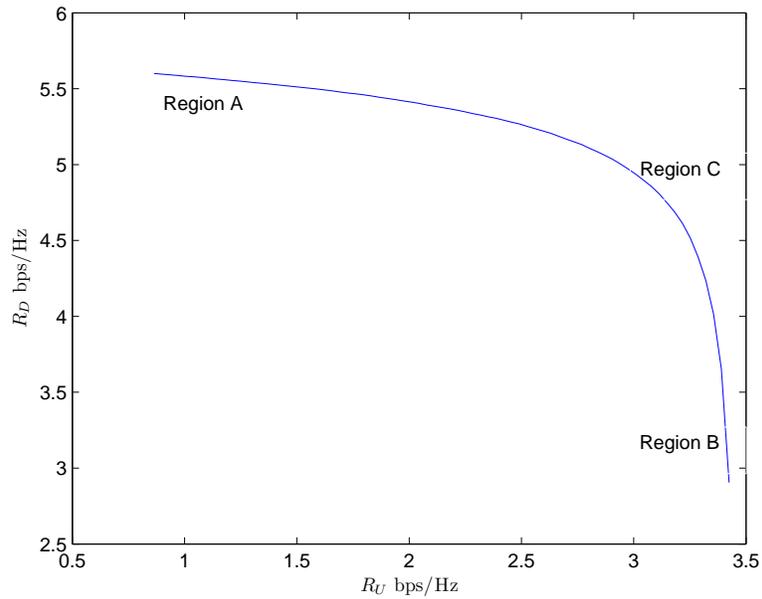}
\caption{A plot of the downlink rate against the uplink rate in a FDD system at 10 dB.}
\label{RdvsRu}
\end{figure}

In dedicated feedback, each user is allocated orthogonal feedback resources per coherence block i.e., a fixed number of symbols on the uplink for SNR feedback. Each user uses $L_{fb}$ bits for SNR feedback. The number of feedback symbols per user will be $\frac{L_{fb}}{\log_2\left( 1+P \right)}$. So the feedback bandwidth in
(\ref{FDD_data_bw}) is $W_{fb}(K)=\frac{KL_{fb}}{\log_2\left( 1+P \right)}=O(K)$ and the uplink rate of a FDD system is
\begin{align}
R_{U}\left( K \right)&=\left(1-\frac{KL_{fb}}{\check{T}\log_2\left( 1+P \right)} \right)\log_2\left( 1+P \right)
=\log_2\left( 1+P \right)-\frac{K}{T}\label{uplinkrate}
\end{align}
where $T=\frac{\check{T}}{L_{fb}}$ is the effective blocklength. 
The downlink rate of a FDD system is $R_{D}\left( K \right)=\mathbb{E}\left[\log_2\left(1+P \gamma_{df}\right)\right]$.\footnote{ We ignore the effect of $L_{fb}$ bit quantization of the SNR value at the user and assume that the feedback channel is error free. These simplifications simplify the analysis yet capturing the cost-benefit tradeoff of multiuser diversity which is primarily quantified in terms of the number of feedback users. These simplications are reasonable enough since 
$L_{fb}$ is large enough such that the high precision quantized SNR value very closely approximates the unquantized SNR value, the errors in the feedback channel can be made very small by using standard QAM methods \cite{Kobayashi}.}
The cost of an additional feedback user is $\frac{dR_{U}}{dK}=\frac{-1}{T}$ and it specifies the decrease in the uplink rate. The benefit of an additional feedback user is $\frac{dR_{D}}{dK}=O\left(\frac{1}{K\log K}\right)$ since $R_{D}\left( K \right)=O\left(\log \log K\right)$ from (\ref{gKapprox}) and it specifies the marginal increase in the downlink rate. The tradeoff between the uplink and downlink rates is more clearly illustrated in Fig.\ref{RdvsRu}. From the figure, it is very clear that an optimized system should operate in region C which achieves a better tradeoff between the uplink and downlink rates than the regions A and B. The optimal operating point on the tradeoff curve can be quantified by solving the optimization problem in (\ref{FDDoptim}),
\begin{equation}
 K^{df}_{op}= \arg \max_K \left[\lambda_R \mathbb{E}\left[\log_2\left(1+P \gamma_{df}\right)\right]+\log_2\left( 1+P \right)-\frac{K}{T}\right]\label{FDDActopt}
\end{equation}
The objective function in (\ref{FDDActopt}) is concave and at the optimal operating point, the benefit of an additional feedback user on the downlink will equal the cost of an additional feedback user on the uplink i.e., 
$\frac{dR_{D}}{dK}=\frac{-1}{\lambda_R}\frac{dR_{U}}{dK}$. It is very difficult to obtain a closed form expression for $K^{df}_{op}$. In an effort to get an expression for $K^{df}_{op}$, we define a tight approximation to the downlink rate based on the upper and lower bounds in (\ref{LUB}), $R_{D}\left( K \right)\approx \tilde{R}_{D}\left( K \right)=\log_2\left( 1+P\log K \right)$ . The optimization problem is defined for the approximation 
\begin{equation}
 K^{df}_{ap}= \arg \max_K \left[\lambda_R \log_2\left( 1+P\log K \right)+\log_2\left( 1+P \right)-\frac{K}{T}\right]\label{FDDApropt}
\end{equation}
The solution to the above optimization problem is given in the following theorem.

\begin{Theoi1}
In a FDD single antenna broadcast channel using dedicated feedback, the number of feedback users $K^{df}_{ap}$ which maximize the approximate weighted sum rate as per (\ref{FDDApropt}) is 
\begin{equation}
K^{df}_{ap}= \frac{\lambda_RT}{\mathcal{W} \left[e^{\frac{1}{P} \lambda_RT}\right] } \approx\frac{\lambda_RT}{\frac{1}{P}+\log
\lambda_RT-\log\left(\frac{1}{P}+\log\lambda_RT\right)}
\label{Kap}
\end{equation} 
where $\mathcal{W}(\cdot)$ is the Lambert-W function.
\end{Theoi1}
\begin{IEEEproof}
 See Appendix \ref{sec:them1proof}.
\end{IEEEproof}

\subsubsection{Scaling of $K_{ap}^{df}$ w.r.t $T$}
\label{KapFDD_scal_T}
From (\ref{Kap}), it is easily seen that $K_{ap}^{df}=O\left(\frac{T}{\log T} \right) $. As $T$ increases, the uplink bandwidth increases and hence an optimized system should increase the number of feedback users. To see why $K_{ap}^{df}$ does not scale linearly with $T$, let us consider how the weighted sum rate in (\ref{FDDApropt}) grows as $T\rightarrow\infty$. The weighted sum rate for $K=O(T)$ and $K=O\left( \frac{T}{\log T} \right)$ are given below
\begin{align}
K=cT, 0<c<1: \lambda_RR_D+R_U \approx \lambda_R \log\log T + \left(1+\lambda_R\right)\log P-c.\label{Linsca} \\
K=O\left( \frac{T}{\log T} \right):\lambda_RR_D+R_U\approx \lambda_R \log\log T + \left(1+\lambda_R\right)\log P-\frac{1}{\log T}.\label{sublinsca}
\end{align}
From (\ref{Linsca}) and (\ref{sublinsca}), it is clear why 
$K_{ap}^{df}\neq O(T)$.

\begin{figure}
\centering
\includegraphics[scale=0.7,width=0.7\textwidth]{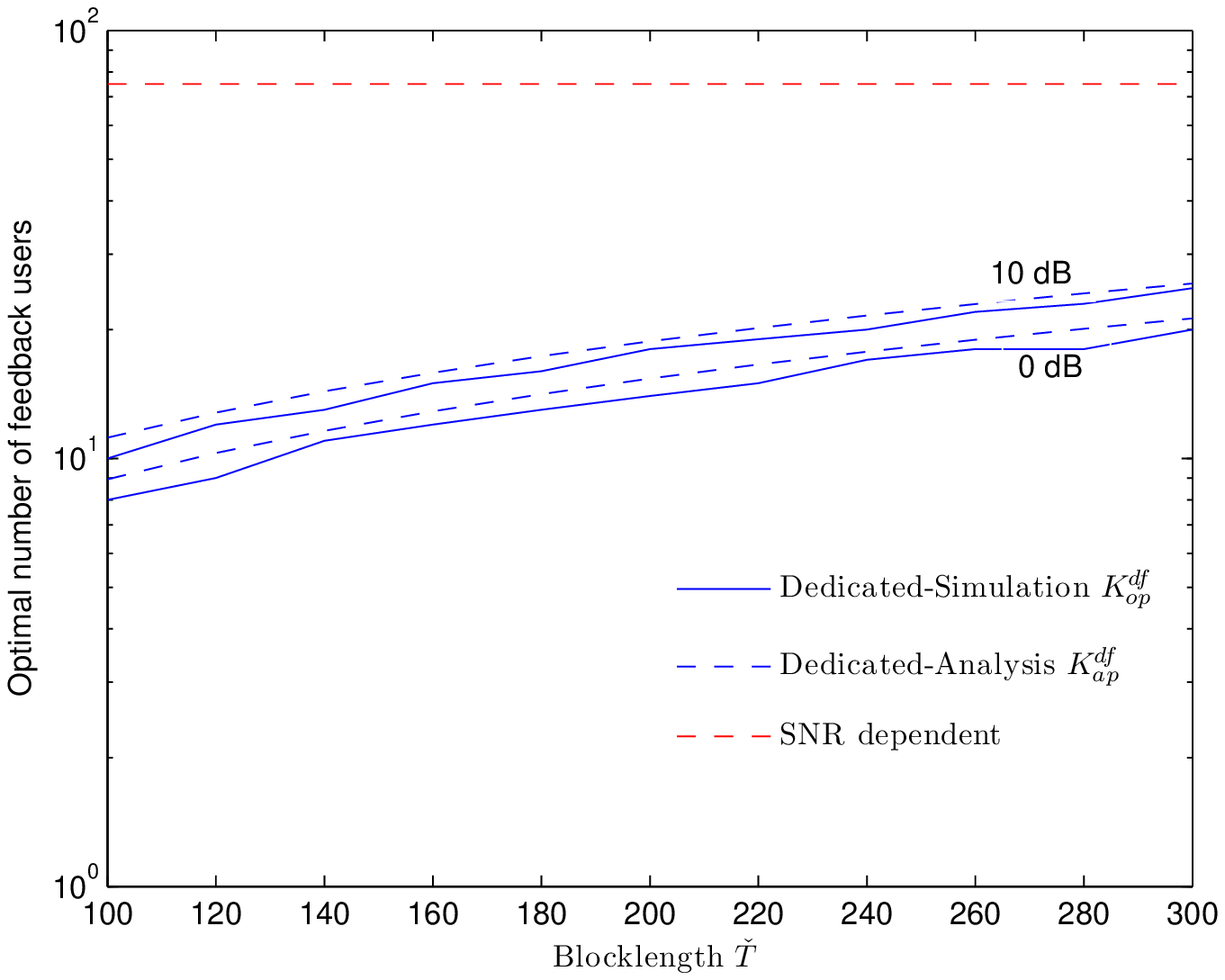}
\caption{Optimal number of feedback users against the blocklength in a FDD system for the 2 SNR feedback methods
.}
\label{KopFDD}
\end{figure}

Before giving the numerical results, the simulation setup is briefly described. The chosen system parameters are $T_c=1$ms, $W_c$ has the range $100-300$kHz which corresponds to the blocklength $\check{T}=100-300$, $L_{fb}=5$ and $\lambda_R=0.5$. The number of users in the system is 75. In Fig.\ref{KopFDD}, $K_{op}^{df}$ and $K_{ap}^{df}$ are plotted against the blocklength at 10 and 0 dB. The scaling of both $K_{op}^{df}$ and $K_{ap}^{df}$ w.r.t $T$ is very similiar. It is seen that for a FDD 75 user single antenna broadcast channel using dedicated feedback, it is wiser to collect SNR only from a few users, more specifically around 8-25 users. For example, $K_{op}^{df}=8$ for $\check{T}=100$ and $0$ dB, $K_{op}^{df}=25$ for $\check{T}=300$ and $10$ dB.
The number of feedback users in an optimized system should be selected very conservatively because the feedback bandwidth in dedicated feedback is $O(K)$, in contrast to the benefit which is $O(\log \log K)$, and this limits the amount of multiuser diversity that can be used.  

\subsubsection{Scaling of $K_{ap}^{df}$ w.r.t $P$}
\begin{figure}
\centering
\includegraphics[scale=0.7,width=0.7\textwidth]{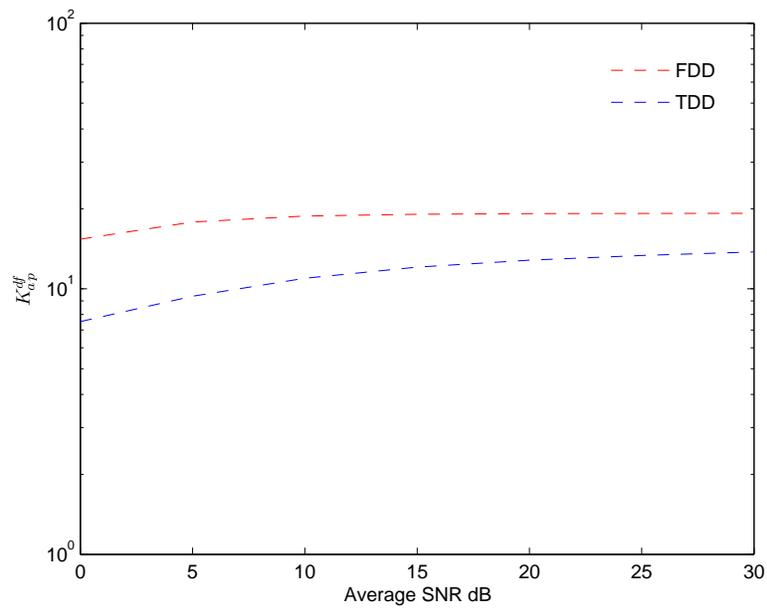}
\caption{$K_{ap}^{df}$ in both the FDD and TDD systems plotted against the average SNR.}
\label{KopvsSNRTDDFDD}
\end{figure}

From (\ref{Kap}), it is easily seen that $K_{ap}^{df}=O(1)$ w.r.t $P$.
Fig.\ref{KopvsSNRTDDFDD} shows a plot of $K_{ap}^{df}$ against $P$. As the average SNR increases, the per user feedback bandwidth decreases. Hence, initially as $P$ increases, the number of feedback users increases in an optimized system, but as $P\rightarrow \infty$, the multiuser diversity benefit diminishes and $O(1)$ scaling is seen.

\subsection{SNR dependent feedback}
\label{CBFB_fdd}
SNR dependent feedback introduced in \cite{Gesbert1} has the basic idea that since only the good users (those with a large instantaneous SNR) will be scheduled, feedback bandwidth can be reduced significantly by asking only the good users to feedback SNR. Users with good SNR are
differentiated from users with bad SNR by a SNR threshold $\gamma_{th}$. The SNR independent feedback method
considered in the previous subsection wastes bandwidth since users with bad SNR also feedback SNR. Users
compare their SNR $\gamma$ to a predetermined $\gamma_{th}$ and only if $\gamma \geq\gamma_{th}$, SNR is fedback on a random access
channel. The choice of $\gamma_{th}$ determines how many users feedback SNR and thus the feedback bandwidth.
The feedback random access channel can be implemented as a slotted ALOHA channel \cite{So,Tang} or CDMA channel \cite{Love}.

We consider the slotted ALOHA like random access channel introduced in \cite{Tang}. The random access channel
has N slots as illustrated in Fig.\ref{Fig2}.c (in this paper, a slot refers to a fixed number of symbols grouped together). A user is eligible to feedback SNR only if $\gamma \geq\gamma_{th}=\log\left( \frac{K}{N} \right)$. An
eligible user will select one of the N slots with probability $\frac{1}{N}$
to feedback its SNR and MS ID. Each
user feeds back $L_{fb}$ bits of SNR value and $\log_2 K$ bits of MS ID. The number of feedback symbols required
for a user is $\frac{L_{fb}+\log_2 K}{\log_2 (1+P )}$.
These many symbols are grouped into one feedback slot. A feedback attempt by
a user is successful (hereafter referred to as user being captured) only if that user is the only one to attempt feedback in its selected slot, because a collision occurs when multiple users select the same slot. After $N$ slots of SNR feedback, the BS selects the user with the
best downlink SNR from amongst the captured users. A more generalized SNR feedback algorithm is
given in \cite{So}, although it is very complex. \\
The uplink data bandwidth is given by
\begin{align}
\mathrm{W_\mathrm{{data}}(K)}&=\left[ 1-\left(\frac{L_{fb}+\log_2K}{\log_2(1+P)}\right)\frac{N}{\check{T}} \right]
\end{align}
The downlink spectral efficiency is 
\begin{align}
C_{sdf}\left( K \right)&=\mathbb{E}[\log_2(1+P\gamma_{sdf})]
\end{align}
$\gamma_{sdf}=\max_{1\leq m \leq X}\gamma_{m\lvert\gamma_{th}}$ is the scheduled SNR, $\gamma_{m\lvert\gamma_{th}}$ is the SNR of the $m^{th}$ captured user and $X$ is the number of captured users. The sum rate is
\begin{align}
    R_D(K)+R_U(K)&=P\left(X\geq 1\right)\left[\mathbb{E}[\log_2(1+P\gamma_{sdf})]+ \left[1-\left(\frac{L_{fb}+\log_2K}{\log_2(1+P)}\right) \frac{N}{\check{T}}\right]\log_2(1+P)\right]\label{CBFB_sumrate}
\end{align}
An expression for $P \left(X \geq 1\right)$ is given in Appendix \ref{CBFB_Proof}. The weighted sum rate optimization is
\begin{align}
    \langle K,N\rangle ^{sdf}_{op}&=\arg \max_{K,N} \left[ \lambda_R R_D(K)+R_U(K)\right]
    \label{CBOptRateFDD}
\end{align}

Based on the expression for the sum rate (\ref{CBFB_sumrate}) in Appendix \ref{CBFB_Proof}, it is easily seen that the optimization in
(\ref{CBOptRateFDD}) does not have a closed form solution and has to be solved numerically. Intuitively, the value of $K^{sdf}_{op}$
can be easily understood. The feedback bandwidth in SNR dependent feedback (\ref{CBFB_sumrate}) is $O(\log K)$, unlike $O(K)$ in dedicated feedback, 
and this literally allows for all of the available multiuser diversity to be used i.e., all users in the
system can participate in the feedback process. In Fig.\ref{KopFDD}, $K^{sdf}_{op}$ is plotted against the blocklength at 10
and 0 dB. It is seen that all the 75 users in the system can participate in the feedback process. 

\begin{figure}
\centering
\includegraphics[scale=0.7,width=0.7\textwidth]{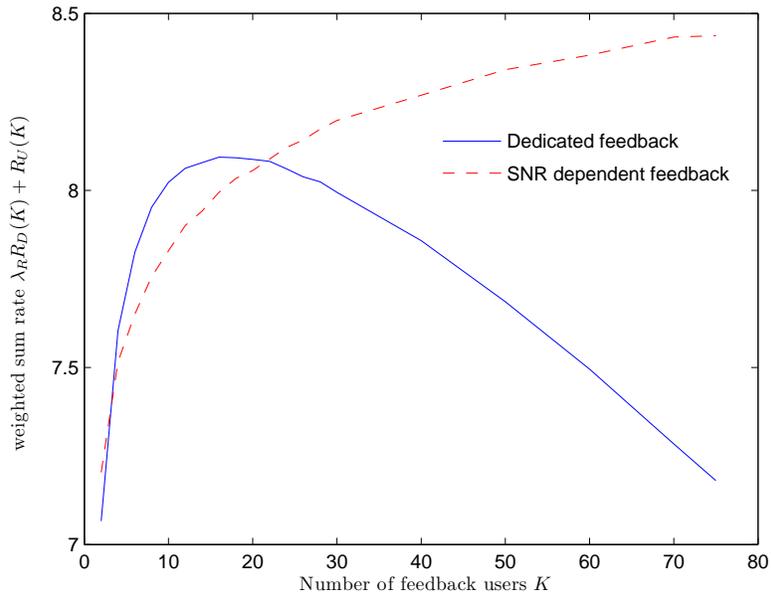}
\caption{Weighted sum rates of both the feedback methods against the number of feedback users in a FDD system
.}
\label{dfandsdf_FDD}
\end{figure}

Fig.\ref{dfandsdf_FDD} plots the weighted sum rates of both the feedback methods against the number of feedback users. Fig.\ref{dfandsdf_FDD} summarizes the key findings of this section i.e., for dedicated feedback, due to the significant feedback cost, the number of feedback users has to be very carefully selected, and for SNR
dependent feedback, due to the relatively much smaller feedback cost, all the users in the system can
participate in the feedback process. 

\section{TDD System}
\label{sec:TDDproblem}
In this section, a cost-benefit analysis of multiuser diversity in a TDD system is performed. We first study dedicated feedback and then address SNR-dependent feedback.

\subsection{Dedicated Feedback}
The feedback process in a TDD system is the same as in a FDD system (section \ref{FDD_ded_fb}) except that it
consumes bandwidth common to both the uplink and downlink. The downlink rate of a TDD system is\footnote{We assume that there is no uplink data in a TDD system. Even if uplink data is present, the weighted sum of the uplink and downlink rates can be rewritten in the same form as (\ref{tdd_dl_rate}). The mathematical details are omitted due to space limitations. Hence assuming that there is no uplink data simplifies the analysis without changing the cost-benefit tradeoff of multiuser diversity.}
\begin{align}
R_{D}\left( K \right)&=\mathrm{W_\mathrm{{data}}(K)} \cdot
C_{df}\left( K \right)=
\left(1-\frac{KL_{fb}}{\check{T}\log_2\left( 1+P \right)} \right)\mathbb{E}\left[\log_2\left(1+P \gamma_{df}\right)\right]
\label{tdd_dl_rate}
\end{align}

\begin{figure}
\centering
\includegraphics[scale=0.7,width=0.7\textwidth]{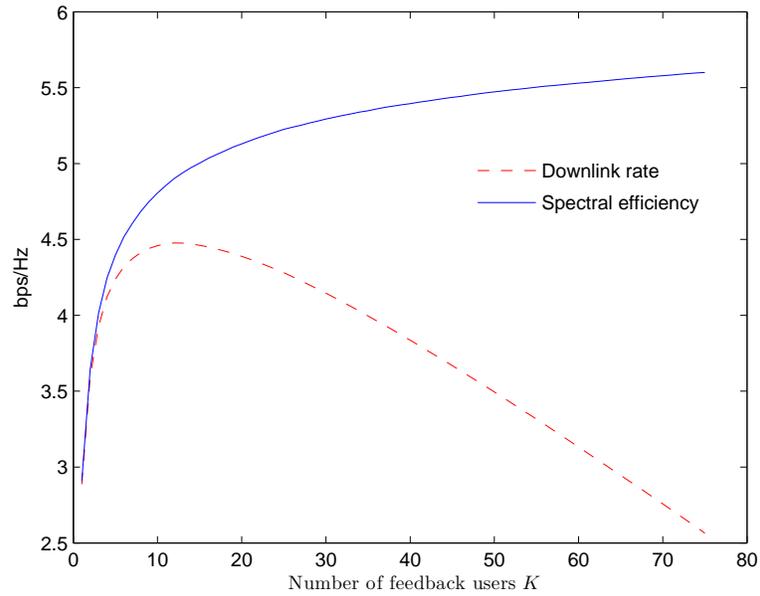}
\caption{Spectral efficiency and downlink rate plotted against the number of feedback
users at 10 dB in a TDD system.}
\label{SEandRd}
\end{figure}

In the TDD system, the cost of an additional feedback user is $\frac{L_{fb}}{\log_2\left( 1+P \right)}$
downlink symbols. The
benefit of an additional feedback user is the increase in spectral efficiency quantified by $\frac{dC_{df}(K)}{dK}=O \left( \frac{1}{K\log K} \right) $. This tradeoff can be better understood from Fig.\ref{SEandRd}, which shows a plot of the spectral
efficiency and the downlink rate against the number of feedback users at 10 dB. Based on this plot, it is very clear that beyond a certain point (the knee region in the spectral efficiency curve), the benefit from
feedback is too small, not worth increasing the feedback cost. So the intuition is to operate around the
knee region achieving an optimal balance between the spectral efficiency benefit and the feedback cost. The knee
region can be quantified by solving the optimization problem
\begin{equation}
 K^{df}_{op}= \arg \max_K \left(1-\frac{K}{T\log_2\left( 1+P \right)} \right)\mathbb{E}\left[\log_2\left(1+P \gamma_{df}\right)\right]\label{TDDActopt}
\end{equation}
It is very difficult to quantify $K^{df}_{op}$ in closed form. In an effort to get an expression for $K^{df}_{op}$, an approximation to the downlink rate based on (\ref{LUB}) is used, $R_{D}\left( K \right)\approx \tilde{R}_{D}\left( K \right)=$ $\left(1-\frac{K}{T\log_2\left( 1+P \right)} \right)$ $\log_2\left( 1+P\log K \right)$. The optimization problem for the approximation is
\begin{equation}
 K^{df}_{ap}= \arg \max_K \left(1-\frac{K}{T\log_2\left( 1+P \right)} \right)\log_2\left( 1+P\log K \right)\label{TDDApropt}
\end{equation}
The objective function is concave and the following theorem quantifies the scaling of $K^{df}_{ap}$ w.r.t $T$.

\begin{Theoi2}
In a TDD single antenna broadcast channel using dedicated feedback, the number of feedback users $K^{df}_{ap}$ which maximize the approximate downlink rate as in (\ref{TDDApropt}) scales with the blocklength $T$ as 
\begin{equation}
K^{df}_{ap}= O \left( \frac{T}{\log T \cdot \log \log T} \right)
\label{Kap_tdd}
\end{equation} 
\end{Theoi2}
\begin{IEEEproof}
 See Appendix \ref{sec:them2proof}, where the following expression for $K^{df}_{ap}$ is derived.
 \begin{equation}
K^{df}_{ap}\approx \frac{T_2}{\mathcal{W} \left[ e^{\frac{1}{P}}T_2\right]},~T_2=\frac{T_1-K_1}{\log \left( 1+P\log K_1 \right)},~K_1= \frac{T_1}{\mathcal{W} \left[ e^{\frac{1}{P}}T_1\right]},~T_1=T\log_2\left( 1+P \right)
\label{Kap_tdd_form}
\end{equation}
\end{IEEEproof}

\begin{figure}
\centering
\includegraphics[scale=0.7,width=0.7\textwidth]{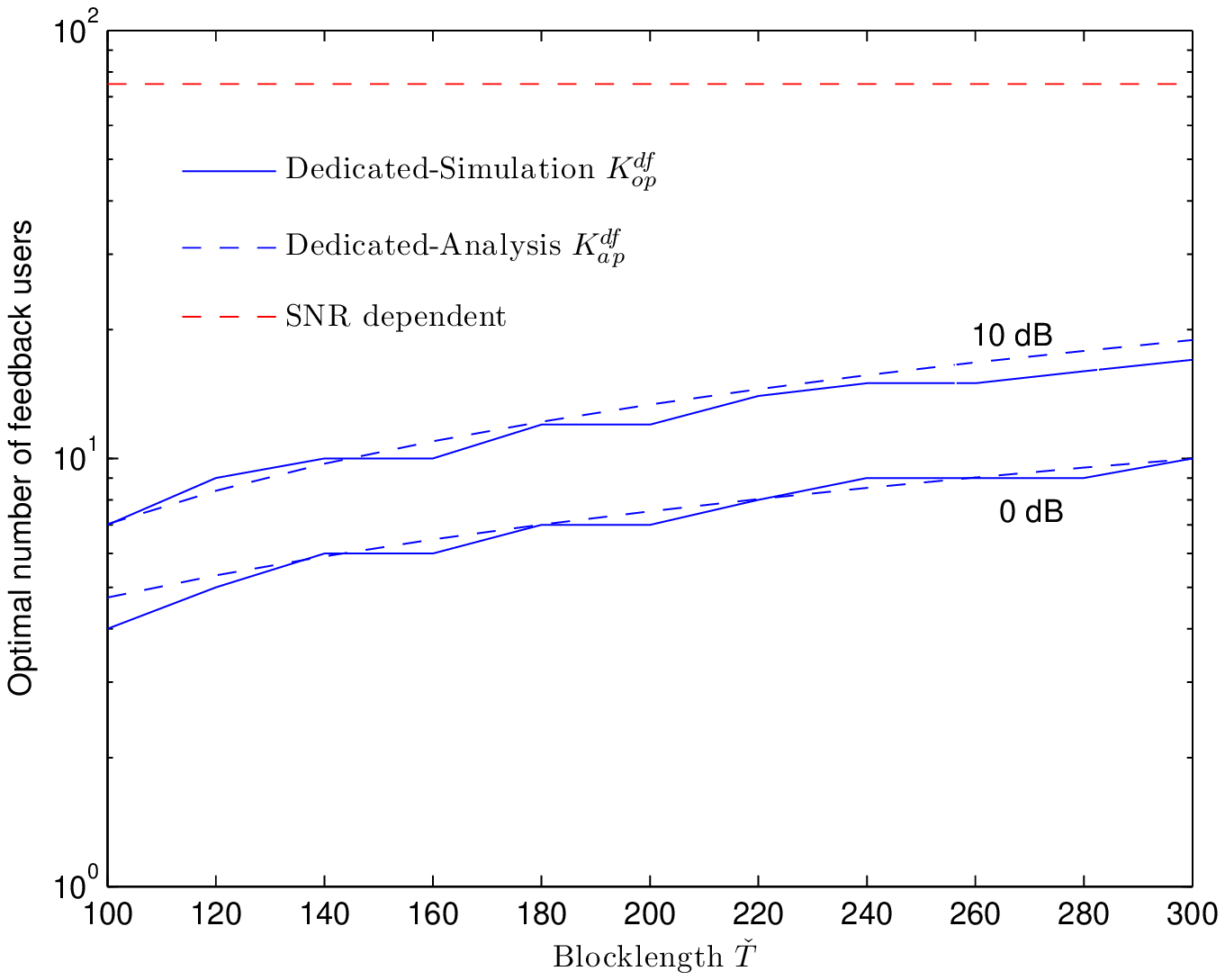}
\caption{Optimal number of feedback users against the blocklength in a TDD system for the 2 SNR feedback methods
.}
\label{KopTDD}
\end{figure}

In Fig.\ref{KopTDD}, $K_{op}^{df}$ and $K_{ap}^{df}$ are plotted against the blocklength at 10 and 0 dB. Similiar to the FDD system, both $K_{op}^{df}$ and $K_{ap}^{df}$ scale w.r.t $T$ is the same way. It is seen that in a TDD 75 user single antenna broadcast channel using dedicated feedback, it is strictly advisable to collect SNR only from a few users around 4-18 users. For example, $K_{op}^{df}=4$ for $\check{T}=100$ and $0$ dB, $K_{op}^{df}=18$ for $\check{T}=300$ and $10$ dB. In a TDD system, the feedback bandwidth affects the downlink rate unlike a FDD system, where the feedback bandwidth affects only the uplink rate and hence the scaling of $K_{ap}^{df}$ in a TDD system (\ref{Kap_tdd}) is more conservative than that of a FDD system in \ref{KapFDD_scal_T}.
\subsubsection{Scaling of $K_{ap}^{df}$ w.r.t $P$}
 It can be seen that $K_{ap}^{df}=O(1)$ after some high SNR analysis in (\ref{Kap_tdd_form}) (mathematical details are omitted due to space limitations). $K_{ap}^{df}$
is plotted against $P$ in Fig.\ref{KopvsSNRTDDFDD} and $O(1)$ scaling is observed. The intuition behind the scaling is the
same as in the FDD section.

\subsection{SNR dependent feedback}
The feedback process is the same as described in the subsection \ref{CBFB_fdd} for the FDD systems except that it takes place in the
common bandwidth in the TDD systems. The downlink rate is given by
\begin{align}
R_D(K)&=P\left(X\geq 1\right)\left[ 1-\left(\frac{L_{fb}+\log_2K}{\log_2(1+P)}\right)\frac{N}{\check{T}} \right]\mathbb{E}[\log_2(1+P\gamma_{sdf})]
\end{align}
 The downlink rate optimization is
\begin{align}
\langle K,N\rangle^{sdf}_{op}&=\arg \max_{K,N} R_D(K)
    \label{CBOptRateTDD}
\end{align}

The optimization in (\ref{CBOptRateTDD}) is difficult to solve in closed form and needs to be solved numerically. In Fig.\ref{KopTDD},
$K^{sdf}_{op}$ is plotted against the blocklength at 10 and 0 dB. Similiar to the FDD system, all the 75 users in the system can
participate in the feedback process. The interpretation of the scaling of $K^{sdf}_{op}$ is the same as in the FDD system.

\begin{figure}
\centering
\includegraphics[scale=0.7,width=0.7\textwidth]{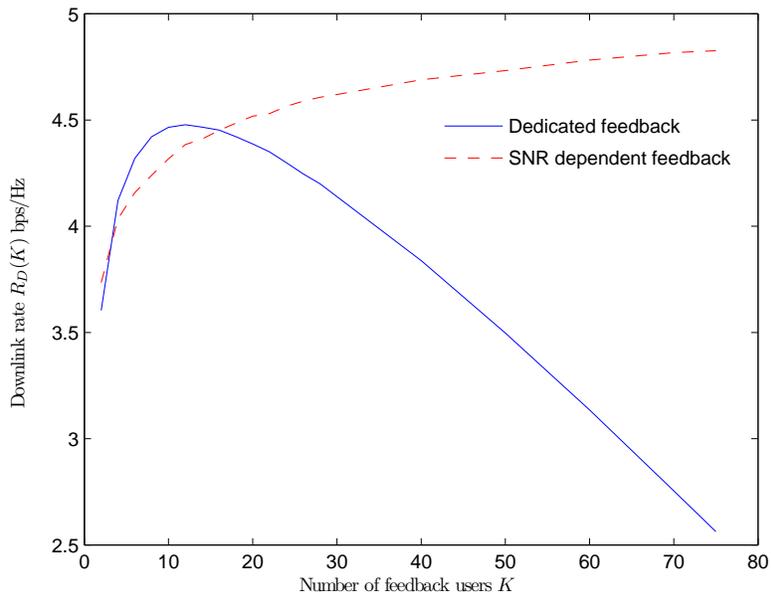}
\caption{Downlink rates of both the feedback methods against the number of feedback users in a TDD system
.}
\label{dfandsdf_TDD}
\end{figure}

The key findings of this section w.r.t the dedicated and SNR dependent feedback methods are the same as in the FDD section and are highlighted in Fig.\ref{dfandsdf_TDD}, which  shows a plot of the downlink rates of both the feedback methods against the number of feedback users.

\section{Single user multiantenna techniques and multiuser diversity}
\label{sec:SU_Multiant}
Multiuser diversity makes use of SNR variation among users. The previous 2 sections quantify the optimal number of feedback users in single antenna broadcast channels using dedicated feedback. But 
when single user multiantenna techniques are employed at BS and/or users, the SNR or mutual information variation tends to cease with the spatial dimension, which suggests that the number of feedback users neccessary in dedicated feedback should decrease with the spatial dimension. 
This effect of single user multiantenna techniques on multiuser diversity in a broadcast channel using dedicated feedback is studied in the next 2 subsections, specifically for MIMO spatial multiplexing and SIMO.

\subsection{MIMO}
\label{MIMO_vblast}
In this subsection, we study a broadcast channel where the BS uses single user MIMO spatial multiplexing, more specifically, V-BLAST. The BS has $N$ Tx antennas and each user has $N$ Rx antennas. The spectral efficiency of $i^{th}$ user  is
\begin{align}
    C_i&=\log_2\Big \lvert \mathbf{I}+\frac{P}{N}\mathbf{H}_i\mathbf{H}_i^H \Big \rvert \label{mimoMI}
\end{align}
where $\mathbf{H}_i=\left[h_{ij}\right]^N_{i,j=1}$ is the $i^{th}$ user $N \times N$ MIMO channel. Users feedback their instantaneous mutual information (M.I) to the BS, which selects the user with the largest M.I for downlink transmission. Assuming a TDD system, the downlink rate is 
\begin{align}
R_{D}(K)&=\left(1-\frac{K}{T\log_2(1+P)} \right)\mathbb{E} \left[ \max_{1\leq i \leq K}\log_2\Big \lvert \mathbf{I}+\frac{P}{N}\mathbf{H}_i\mathbf{H}_i^H \Big \rvert \right] \label{mimoDLrate}
\end{align}
In an effort towards getting an expression for the optimal number of feedback users, $K_{op}^{df}=\arg \max_KR_{D}(K)$, the following Gaussian approximation to the mutual information (\ref{mimoMI}) is used \cite{Hochwald}.
\begin{align}
&\log_2\Big \lvert \mathbf{I}+\frac{P}{N}\mathbf{H}_i\mathbf{H}_i^H \Big \rvert \sim \mathcal{N}(\mu,\sigma^2)\nonumber\\
&\mu=N\log_2 \left(\frac{P}{e} \right) ,~\sigma^2= \left(\log_2e\right)^2 [\log N+1.58]\label{mimoMIap}
\end{align}
Using a standard result from Order Statistics \cite{OSbook2}, the downlink rate in (\ref{mimoDLrate}) is approximated as
\begin{align}
R_{D}(K)&\approx \tilde{R}_{D}(K)= \left(1-\frac{K}{T\log_2(1+P)} \right)\left[\mu+\sigma \sqrt{2\log K}\right] \label{mimoDLrateap}
\end{align}

\begin{figure}
\centering
\includegraphics[scale=0.7,width=0.7\textwidth]{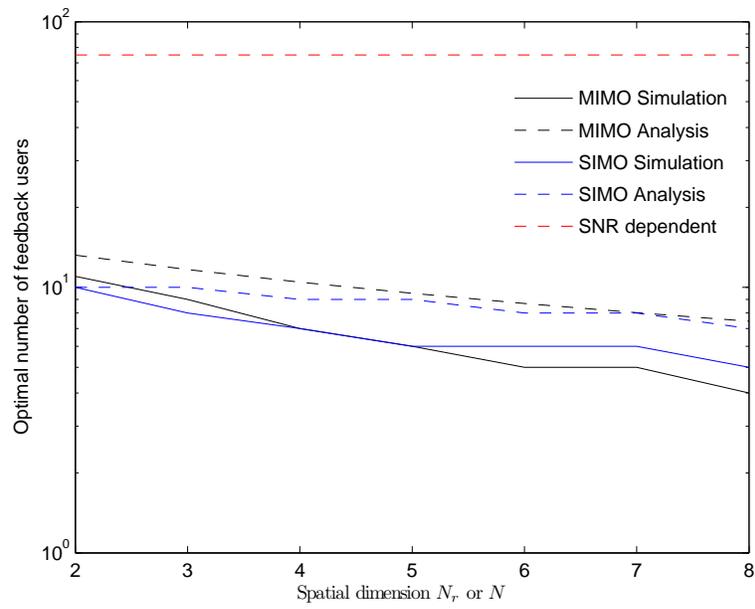}
\caption{Optimal number of feedback users against the spatial dimension in a broadcast channel using single user multiantenna techniques and dedicated feedback.}
\label{KopMIMO_SIMO}
\end{figure}

The downlink rate approximation in (\ref{mimoDLrateap}) is concave and thus, the approximation to the optimal number of feedback users, $K_{ap}^{df}=\arg \max_K\tilde{R}_{D}(K)$, is easily given by
\begin{align}
K_{ap}^{df}& \approx\frac{T\log_2(1+P)}{2\mathcal{W}\left[\frac{BT\log_2(1+P)}{2}\right]},~\log B=0.5\left[1+\frac{\mu}{\sigma}\sqrt{2\log K_1}\right],~K_1=\frac{T\log_2(1+P)}{2\mathcal{W}\left[\frac{T\log_2(1+P)}{2}\right]} \label{mimoapKap}
\end{align}
It is easily seen from (\ref{mimoapKap}) that $K_{ap}^{df}=O\left(\frac{\sqrt{\log N}}{N}\right)$. Fig.\ref{KopMIMO_SIMO} shows a plot of both $K_{op}^{df}$ and $K_{ap}^{df}$ against $N$. It is seen that both $K_{op}^{df}$ and $K_{ap}^{df}$ decrease with $N$.

\subsection{SIMO}
In this subsection, we study the effect of multiple receiver antennas at the users on multiuser diversity. Downlink is SIMO and uplink is SISO. Each user has $N_r$ receiver antennas. The $i^{th}$ user downlink channel is $\vec{\textbf{h}}_i= \left[h_{i1},h_{i2},\cdots,h_{iN_r}\right]$ and the downlink SNR is $\gamma_i=\sum_{j=1}^{N_r}\abs{h_{ij}}^2$. The spectral efficiency is given by $C_{df}(K)=\mathbb{E}[\log_2(1+P\gamma_{df})]$, where $\gamma_{df}=\max_{1\le i \le K}\gamma_i$. The downlink rate is $ R_{D}(K)=\left(1-\frac{K}{T\log_2(1+P)} \right)\mathbb{E}[\log_2(1+P\gamma_{df})]$. The optimal number of feedback users is $K_{op}^{df}=\arg \max_KR_{D}(K)$. In an effort towards getting an expression for $K_{op}^{df}$, the following approximate expression for the downlink rate is used.
\begin{align}
R_{D}(K)\approx \tilde{R}_{D}(K)=\left(1-\frac{K}{T\log_2(1+P)} \right)\log_2(1+Pb_K) \label{SIMODLrateap}
\end{align}
where $b_K$ is the growth rate of $\gamma_{df}$. An expression for $b_K$ is given in Appendix \ref{sec:SIMO_proof}.

An approximation to the optimal number of feedback users is $K_{ap}^{df}=\arg \max_K \tilde{R}_{D}(K)$. 
In Fig.\ref{KopMIMO_SIMO}, a plot of $K_{op}^{df}$ and $K_{ap}^{df}$ against $N_r$ is shown.  Although an approximate expression for $K_{ap}^{df}$ can be written out, it is too complex and hence, $K_{ap}^{df}$ was solved numerically. Similiar to the MIMO case, both $K_{op}^{df}$ and $K_{ap}^{df}$ decrease with $N_r$.
 
The key finding of this section is that a broadcast channel using single user multiantenna techniques and dedicated feedback should decrease the multiuser diversity order with the spatial dimension, which agrees with the findings in \cite{Kobayashi}. 
This take away message is not applicable to a broadcast channel using single user multiantenna techniques and SNR or MI dependent feedback. This is true since in the SNR or MI dependent feedback, all the users in the system can participate in the feedback process, although the number of slots $N$ in the random access channel should decrease with the spatial dimension.
\section{Conclusion}
\label{conc_sec}
The cost-benefit analysis of multiuser diversity in single antenna broadcast channels using both
dedicated and SNR dependent feedback methods was performed.
A single antenna broadcast channel using dedicated feedback due to the significant feedback cost has to very carefully select the number of feedback users. 
A single antenna broadcast channel using SNR
dependent feedback due to the relatively much smaller feedback cost can have all the users in the system participate in the feedback process i.e., all of the
available multiuser diversity can be used.
The effect of single user multiantenna techniques on the multiuser diversity order in a broadcast channel using dedicated feedback was studied. A broadcast channel using single user multiantenna techniques and dedicated feedback should decrease the multiuser diversity order with the spatial dimension, since the single user multiantenna techniques decrease 
the SNR or MI variation with the spatial dimension.

\appendices
\section{Proof of Proposition 1}
\label{sec:prop1proof}
Jensen's upper bound for $C_{df}(K)$ is $E\left[\log_2
\left(1+P \gamma_{df}\right)\right] \le \log_2 \left(1+P E\gamma_{df}\right)$. From \cite{OSbook2}, it is known that
$E\gamma_{df}=\sum_{i=1}^{K}\frac{1}{i}$. From \cite{TableBook}, $\sum_{i=1}^{K}\frac{1}{i}=0.58 + \log K$. Thus, $C_{df}(K)$ is upper bounded as $C_{df}(K) \le \log_2 \left(1+P\left(0.58 + \log K\right)\right)$.
Applying Markov Inequality (For a non-negative RV $X$, $E[X]\geq P(X\geq \delta)\delta$, for any $\delta>0$) to $C_{df}(K)$, we get
\begin{align*}
&E\left[\log_2 \left(1+P \gamma_{df}\right)\right] \geq \log_2 \left(1+P \delta\right)P\left(\gamma_{df}\geq \delta\right).
\end{align*}
Picking $P\left(\gamma_{df}\geq \delta\right)=1-\frac{1}{\left(\log K\right)^s},~s>0$,
a corresponding value of $\delta$ can be choosen
\begin{align*}
&P\left(\gamma_{df}\geq \delta\right)=1-\frac{1}{\left(\log K\right)^s}=1-\left(1-e^{-\delta}\right)^K
\Rightarrow
\delta=\log K-\log \left(s\log \log K\right)
\end{align*}
Letting $s=1$, we get a lower bound for $C_{df}(K)$
\begin{align*}
C_{df}(K) &\geq \left(1-\frac{1}{\log
K}\right)\log_2 \left(1+P \left(\log K-\log\log\log K\right)\right)\\
&= \log_2 \left(1+P \left(\log K-\log\log\log K\right)\right)+ o(1)
\end{align*}
Thus $C_{df}(K)$ is bounded as
\begin{align*}
&\log_2 \left(1+P \left(\log K-\log\log\log K\right)\right)+ o(1)\leq
C_{df}(K) \leq \log_2 \left(1+P\left(0.58 +
\log  K\right)\right)
\end{align*}

\section{Proof of Theorem 1}
\label{sec:them1proof}
The objective function in (\ref{FDDApropt}) is concave. $K^{df}_{ap}$ is easily found by setting the derivative to zero.
\begin{align*}
&\frac{\lambda_RP}{K\left( 1+P\log K \right)}-\frac{1}{T}=0\\
&K\left(\frac{1}{P} +\log K \right)=\lambda_RT \Rightarrow K\log BK=\lambda_RT\\
&K^{df}_{ap}=\frac{\lambda_RT}{\mathcal{W} \left[e^{\frac{1}{P} \lambda_RT}\right] } \approx\frac{\lambda_RT}{\frac{1}{P}+\log
\lambda_RT-\log\left(\frac{1}{P}+\log\lambda_RT\right)}
\end{align*}
\section{Proof of $P \left(X \geq 1\right)$ in (\ref{CBFB_sumrate})}
\label{CBFB_Proof}
$X$ represents the number of captured SNR's. Let $L$ represent the number of SNR's above the threshold. It is easily seen that $L\sim \mathrm{Binomial}\left(K,p\right) $, where $p=\frac{N}{K} $ since $\gamma_{th}=\log\left( \frac{K}{N} \right)$. An expression for $P \left(X \geq 1\right)$ is derived below.
\begin{align}
P \left(X \geq 1\right)&=1-P \left(X = 0\right)=1-\sum_{l=1}^KP \left(X = 0\lvert L=l\right)P \left(L=l\right)\nonumber\\
&=1-\sum_{l=1}^KP \left(X = 0\lvert L=l\right) \binom{K}{l} p^l\left(1-p\right)^{K-l}\nonumber\\
&=1-\sum_{l=1}^KP \left(X = 0\lvert L=l\right) \binom{K}{l} \frac{N^l}{K^K} \left(K-N\right)^{K-l}
\end{align}
$P \left(X = 0\lvert L=l\right)$ represents the probability that each of the $N$ slots is accessed by multiple feedback users $(>1)$. 
\begin{align}
&P \left(X = 0\lvert L=l\right) =
\frac{1}{l^N}\left[N+\sum_{i=2}^{\lfloor\frac{l}{2}\rfloor}
N\left(N-1\right)\cdots\left(N-\left(i-1\right)\right) ~!l\sum_{j=1}^{\lvert S_i \rvert }\Phi \left(S_i^j\right)\prod_{k=1}^i \frac{1}{!S_i^j(k)}\right]
\end{align}
where\\
$S_i$: Set of all $i$-tuples s.t all elements of a tuple add upto $N$,\\
$S_i^j$: $j^{th}~ i-$tuple in $S_i$,\\
$S_i^j(k)$: $k^{th}$ element of $S_i^j$ and $\sum_{k=1}^iS_i^j(k)=N$,\\
$\Phi \left(S_i^j\right)=\frac{1}{!m_1!m_2\cdots !m_i}$ where $m_j=[1,i],1 \leq j \leq i$ represents the number of repetitions of an element in the $i$-tuple.

The distribution of $X$ can be written down
completely, but is omitted here due to space limitations. The distribution of $\gamma_{sdf}$ in (\ref{CBFB_sumrate})) is difficult to
write out.
\section{Proof of Theorem 2}
\label{sec:them2proof}
The objective function in (\ref{TDDApropt}) is concave.
\begin{align}
&\tilde{R}'_{D}\left( K \right)=\frac{\frac{1}{K}-\frac{1}{T_1}}{\frac{1}{P} +\log K}-\frac{\log\left( 1+P\log K \right)}{T_1}=0\nonumber\\
&\frac{T_1}{K}-1=\log\left( 1+P\log K \right) \log bK \label{Kap_equat}
\end{align}
where $T_1=T\log_2\left( 1+P\right)$ and $\log b=\frac{1}{P}$.

A $I^{st}$ order solution to (\ref{Kap_equat}) is obtained as
\begin{align}
K\log bK=T_1 \Rightarrow K_1=\frac{T_1}{\mathcal{W} \left(b T_1\right)}
\end{align}
Using $K_1$ in (\ref{Kap_equat}), a more refined solution is obtained.
\begin{align}
K\log bK=T_2=\frac{T_1-K_1}{\log\left( 1+P\log K_1 \right)}\\
\Rightarrow K^{df}_{ap} \approx \frac{T_2}{\mathcal{W} \left(b T_2\right)}=O\left( \frac{T}{\log T\cdot \log \log T } \right)
\end{align}

\section{Proof of (\ref{mimoapKap})}
\label{MIMO_proof}
The approximation to the optimal number of feedback users, as defined in subsection \ref{MIMO_vblast}, is given by
\begin{align}
&K_{ap}^{df}=\arg \max_K \left(1-\frac{K}{T\log_2(1+P)} \right)\left[\mu+\sigma \sqrt{2\log K}\right]
\end{align}
The objective function is concave. 
\begin{align}
& \left(1-\frac{K}{T\log_2(1+P)} \right) \frac{\sigma}{K\sqrt{2\log K}}-\frac{\left[\mu+\sigma \sqrt{2\log K}\right]}{T\log_2(1+P)}=0\nonumber\\
&K\left[1+\frac{\mu}{\sigma}\sqrt{2\log K}+2\log K\right]=T\log_2(1+P) \label{mimo_sol_equ}
\end{align}
$I^{st}$ order solution to (\ref{mimo_sol_equ}) is given by
\begin{align}
K_1=\frac{T\log_2\left( 1+P\right)}{2\mathcal{W} \left[\frac{T\log_2\left( 1+P\right)}{2}\right]}\label{first_order_mimo}
\end{align}
Using (\ref{first_order_mimo}) in (\ref{mimo_sol_equ}), second order solution is obtained.
\begin{align}
K\log BK=\frac{T\log_2\left( 1+P\right)}{2}, \Rightarrow 
K_{ap}^{df}=\frac{T\log_2\left( 1+P\right)}{2\mathcal{W} \left[\frac{BT\log_2\left( 1+P\right)}{2}\right]}
\end{align}
where $\log B=0.5\left[1+\frac{\mu}{\sigma}\sqrt{2\log K_1}\right]$.
\section{Proof of (\ref{SIMODLrateap})}
\label{sec:SIMO_proof}
$\gamma_{df}=\max_{1\le i \le K} \lVert\vec{\textbf{h}}_i\rVert^2 $,
where $\lVert\vec{\textbf{h}}_i\rVert^2 \sim \Gamma\left(N_r,1\right)$.

For the proof of (\ref{SIMODLrateap}), the following lemma from \cite{OSbook2} is used.
\begin{Lem}\label{Lemma1}
  If $X_1,X_2,\cdots,X_K$ are a sequence of positive IID RV's having absolutely continuous cdf
$F_{X}\left(x\right)$ with $\Omega\left(F\right)=\sup\left\{x:F\left(x\right)<1\right\}$ \& there is a real
number $x_1$ such that for all $x_1<x<\Omega\left(F\right)$, $f_{X}\left(x\right)=F_{X}'\left(x\right)$,
$F_{X}''\left(x\right)$ exist \& $f_{X}\left(x\right)\neq0$.
If $\displaystyle{g\left(x\right)=\frac{1-F_{X}\left(x\right)}{f_{X}\left(x\right)}}$ \&  $\lim_{x\to
\Omega\left(F\right)}g\left(x\right)=c,~c\geq0$, then $\displaystyle{\frac{\max_{1\leq i \leq
K}X_{i}-b_K}{a_K}}\stackrel{d}{\rightarrow}e^{-e^{-x}}$, where $b_K=F^{-1}\left(1-\frac{1}{K}\right)$ \&
$a_K=\frac{1}{Kf\left(b_K\right)}$.
Since $g\left(b_K\right)=\frac{1}{Kf\left(b_K\right)}=a_K$, if $\lim_{x\to
\Omega\left(F\right)}g\left(x\right)=c$, then $\lim_{K\to\infty}a_K=c$.
\begin{equation}
        \text{If}~c=0,~\text{then}~\max_{1\leq i \leq K}X_{i}-b_K\stackrel{P}{\rightarrow}0;~
        \text{otherwise}~\frac{\max_{1\leq i \leq K}X_{i}}{b_K}\stackrel{P}{\rightarrow}1
\end{equation}
\end{Lem}

Using the above lemma, the growth rate of $\gamma_{df}$ can be obtained.
The distribution of the IID RV's in
this case is $\Gamma\left(N_r,1\right)$. The $g\left( \gamma \right)$ for this distribution as defined in the
Lemma \ref{Lemma1} is
\begin{align*}
 g\left( \gamma \right)&=\frac{1-F(\gamma)}{f(\gamma)}=
\dfrac{\Gamma\left(N_r\right)e^{-\gamma}\sum_{i=0}^{N_r-1}\frac{\left(\gamma\right)^{i}}{i!}}{\gamma^{N_r-1}e^{
-\gamma}}\\
\lim_{\gamma \to \Omega\left(F\right)=\infty}g\left( \gamma \right)&=\lim_{\gamma \to \infty}
\Gamma\left(N_r\right)\frac{1}{\gamma^{N_r-1}}\sum_{i=0}^{N_r-1}\frac{\left(\gamma\right)^{i}}{i!}=1
\end{align*}
Since $\lim_{\gamma \to \infty}g\left( \gamma \right)=1$, $\frac{\gamma_{df}}{b_K}
\stackrel{P}{\rightarrow}1$.
An expression for $b_K$ is to be obtained in terms of $K$ \& $N_r$ by using the Lemma \ref{Lemma1}.
\begin{align}
        F\left(b_K\right)&=1-\frac{1}{K}=1-e^{-b_K}\sum_{i=0}^{N_r-1}\frac{\left(b_K\right)^{i}}{i!}\nonumber\\
        b_K&=\log K+\log \left(\sum_{i=0}^{N_r-1}\frac{\left(b_K\right)^{i}}{i!}\right)\label{GR}
\end{align}
The above equation is to be solved for $b_K$. The first order approximation is derived below.
\begin{align}
&b_K=\log K+\log \frac{\left(b_K\right)^{N_r-1}}{N_r-1!}\Rightarrow -\left(N_r-1\right)\log b_K+ b_K=\log
\frac{K}{N_r-1!}\label{FOE}
\end{align}
The equation $A\log x+Bx=C$ has the solution $\displaystyle{x=\frac{A}{B}\mathcal{W}\left(\frac{B}{A}
e^{\frac{C}{A}}\right)}$.
Using this solution for $\left(\ref{FOE}\right)$, the first order approximation for $b_K$ is
\begin{align}
b_K&=-\left(N_r-1\right)\mathcal{W}\left[\frac{-1}{\left(N_r-1\right)e^{\frac{1}{N_r-1}\log
\frac{K}{N_r-1!}}}\right]\nonumber\\
b_K&\stackrel{\left(a\right)}{=}-\left(N_r-1\right)\left[-\log\left(\left(N_r-1\right)e^{\frac{1}{N_r-1}\log
\frac{K}{N_r-1!}}\right)-\log \log\left(\left(N_r-1\right)e^{\frac{1}{N_r-1}\log
\frac{K}{N_r-1!}}\right)\right]\nonumber\\
b_K&=\log K + \left(N_r-1\right)\log \left[\log K+\log
\frac{\left(N_r-1\right)^{N_r-1}}{N_r-1!}\right]-\log \left(N_r-1\right)!\label{FOS}
\end{align}
where $\left(a\right)$ follows from a $2^{nd}$ order approximation to $\mathcal{W}\left(x\right)$ \cite{Lambert},
\begin{align*}
\mathcal{W}_{-1}\left(-x\right)&=\log x-\log\left(-\log x\right)+O\left(\frac{\log\left(-\log x\right)}{\log
x}\right),~0 < x\leq \frac{1}{e}.
\end{align*}
 Using
$\left(\ref{FOS}\right)$ in (\ref{GR}), the growth rate of $\gamma_{df}$ is obtained
\begin{align}
        &b_K=\log K +\log \left[ \sum_{j=0}^{N_r-1}\frac{\left(c_K\right)^j}{j!}\right]=O\left(N_r \right) \label{SNRgr2} \\
 &c_K=\log K +\left(N_r-1\right)\log \left[\log K+\log \frac{\left(N_r-1\right)^{N_r-1}}{N_r-1!}\right]-
\log \left(N_r-1\right)!\label{ck2}
\end{align}
%
\bibliography{References}
\bibliographystyle{IEEEtran}

\end{document}